\pdfoutput=1

\documentclass[%
 prr,
 amsmath,amssymb,
 reprint,%
 twocolumn,
 superscriptaddress,
]{revtex4-2}
\usepackage{lipsum}
\usepackage[english]{babel} % nice line breaking of words
\usepackage[normalem]{ulem}
\usepackage{filecontents}
\usepackage[version=3]{mhchem}
\usepackage{color} % Color text package
\usepackage[dvipsnames]{xcolor} % extra colors
\usepackage{siunitx}

\usepackage{multirow}
\usepackage{comment}
\usepackage{makecell}
\usepackage{subcaption}
\usepackage{relsize}

\usepackage{tikz}
\usepackage{mathtools}
\usetikzlibrary{arrows.meta, decorations.pathreplacing,decorations.markings, decorations.pathmorphing, positioning, shapes, calc,fadings}
\tikzfading %strangely gives bad bounding box when inside the tikzpicture
[
name=fade out,
inner color=transparent!0,
outer color=transparent!100
]
\tikzset{snake it/.style={decorate, decoration=snake}}
\usepackage{tikz-feynhand}

\usepackage{graphicx}% Include figure files
\usepackage{dcolumn}% Align table columns on decimal point
\usepackage{bm}% bold math
\usepackage[utf8]{inputenc}
\usepackage[T1]{fontenc}

\usepackage{mathptmx}
\usepackage{mathrsfs}

\DeclareSymbolFont{newfont}{OML}{cmm}{m}{it}% Computer Modern math font
\DeclareMathSymbol{\epsilon}{3}{newfont}{15}% Symbol 15

\usepackage{booktabs}

\usepackage{xr}
\makeatletter
\newcommand*{\addFileDependency}[1]{% argument=file name and extension
  \typeout{(#1)}
  \@addtofilelist{#1}
  \IfFileExists{#1}{}{\typeout{No file #1.}}
}
\makeatother

\newcommand*{\myexternaldocument}[1]{%
    \externaldocument{#1}%
    \addFileDependency{#1.tex}%
    \addFileDependency{#1.aux}%
}
\myexternaldocument{./__bsupp}
\usepackage{prettyref}
\newrefformat{supp-fig}{S\ref{#1}}
\newrefformat{supp-tbl}{S\ref{#1}}
\newrefformat{supp-eq}{S\ref{#1}}
\DeclareUnicodeCharacter{2212}{-}

\usepackage{lipsum}
\usepackage{soul}

\begin{document}

\preprint{}

\title[]{Constraint-aware functional cloning for stable and transferable machine-learned density functional theory}
\newcommand{\ICNDOS}{Catalan Institute of Nanoscience and Nanotechnology (ICN2), Bellaterra, Barcelona  08193, Spain}
\newcommand{\SBUphysast}{Physics and Astronomy Department, Stony Brook University, Stony Brook, New York 11794-3800, United States}
\newcommand{\IACS}{Institute for Advanced Computational Science, Stony Brook University, Stony Brook, New York 11794-3800, United States}

\newcommand{\NYAlec}{The New York Academy of Sciences, New York, New York 10006, United States}
\newcommand{\CFM}{Centro de F\'{i}sica de Materiales (CFM-MPC), CSIC-UPV/EHU, Donostia-San Sebasti\'{a}n 20018, Spain}
\newcommand{\IrvineCh}{Department of Chemistry, University of California, Irvine, CA 92697, USA}
\newcommand{\IrvinePh}{Department of Physics and Astronomy, University of California, Irvine, CA 92697, USA}

\author{Sara Navarro-Rodr\'{i}guez}
\email{sara.navarro@icn2.cat}
\affiliation{\ICNDOS}
%\author{team NN-XC}
\author{Alec Wills}
\affiliation{\SBUphysast}
\affiliation{\IACS}
\affiliation{\NYAlec}
\author{Kimberly J. Daas}
\affiliation{\IrvineCh}
\author{Mar\'{i}a Camarasa-G\'{o}mez}
\affiliation{\CFM}
\author{Marivi Fern\'{a}ndez-Serra}%
\email{marivi.fernandez-serra@stonybrook.edu}
\affiliation{\SBUphysast}
\affiliation{\IACS}

\date{\today}% It is always \today, today,

\begin{abstract}
We study a simple but useful test for neural exchange--correlation (XC) functionals: can a neural model reproduce an established XC functional when it is used self-consistently?
We call this test functional cloning.
The model is trained at the GGA level to reproduce a known semilocal functional, using either a constrained or an unconstrained architecture.
The motivation is that an XC functional is not used on a fixed input.
In a Kohn--Sham self-consistent-field calculation it contributes to the potential, and the resulting density is part of the outcome of the same calculation.
A good pointwise fit to sampled density descriptors is therefore not by itself enough.
Because the target functional is known, the error can be measured directly.
We compare the clones on sampled descriptors, molecular total energies, energy differences, transfer between \textsc{PySCF} and \textsc{siesta}, and equations of state for crystalline solids.
The constrained models reproduce the reference functional more accurately in molecular self-consistent calculations.
They also give better initial parameters for later optimization against correlated molecular energies.
An additional observation is that the constrained architecture already gives a reasonable solid-state baseline before cloning, as seen from randomly initialized constrained models.
Clones trained only on molecular densities transfer well to solids, reproducing reference lattice constants and bulk moduli across metallic, covalent, ionic, oxide, and layered systems.
Cross-code tests show that energy differences are relatively robust, while total energies depend strongly on whether the cloning descriptors come from all-electron or pseudopotential densities.
These results make functional cloning a useful diagnostic before full self-consistent training of neural XC functionals.
\end{abstract}
\maketitle

\section{\label{sec:intro} Introduction}

Density functional theory (DFT) is the workhorse of electronic-structure calculations in condensed-matter physics, chemistry, and materials science because it provides a practical balance between accuracy and computational cost~\cite{hohenberg_inhomogeneous_1964,kohn_self-consistent_1965}.
Its accuracy, however, is limited by the approximation chosen for the exchange--correlation (XC) functional~\cite{cohen_insights_2008}.
In general, semi-local functionals within the generalized-gradient approximation (GGA) such as the Perdew--Burke--Ernzerhof (PBE) functional, remain the default choice in many applications because of their robustness and broad applicability~\cite{perdew_generalized_1996,perdew_jacobs_2001}.
However, their well known limitations continue to motivate the search for improved XC approximations that preserve these practical advantages while increasing accuracy across chemically and structurally diverse systems~\cite{cohen_challenges_2012, becke_perspective_2014, verma_status_2020, horton_promises_2021, teale_dft_2022}.

A rapidly developing route toward this goal is the use of flexible, data-driven parameterizations, including neural networks (NNs), to construct machine-learned XC functionals~\cite{dick_machine_2020, pederson_machine_2022, kulik_roadmap_2022, bystrom_nonlocal_2024, bystrom_cider_2022, luise_accurate_2025}.
These approaches can incorporate complex local and semilocal density representations and can be trained against accurate reference data~\cite{akashi_can_2025}.
There has been a significant amount of progress in the last years, enabled by the development of large benchmark datasets derived from high-level wavefunction methods~\cite{goerigk_look_2017,gould_slim_2025}, differentiable implementations of the self-consistent Kohn--Sham equations~\cite{zhang_differentiable_2022,dick_highly_2021,strachwitz_data-efficient_2026}, and large-scale interdisciplinary efforts, including recent industry-led developments~\cite{kirkpatrick_pushing_2021,luise_accurate_2025}.
These advances have shown that machine learning can fit accurate energies.
The question now is how to design learned functionals that remain stable, physically meaningful, and transferable when embedded inside the self-consistent electronic-structure problem.

This distinction is crucial for scientific machine learning in DFT.
In self-consistent calculations, the functional is not evaluated on a fixed input density; it helps generate the density on which it is then evaluated.
The learned model is therefore part of a nonlinear fixed-point map: changing its parameters changes the XC potential, the density, and the solution reached by the Kohn--Sham equations.
This feedback makes NN-based XC optimization qualitatively different from ordinary supervised learning.
A model that appears accurate on fixed descriptors can still produce unstable self-consistent iterations or produce numerically inconsistent gradients during end-to-end optimization.
Consequently, trainable XC strategies can benefit from an initialization stage that anchors the neural representation to a physically reasonable functional form before full optimization~\cite{dick_highly_2021}.
Such initialization is not merely a numerical convenience; it places the learned model near a meaningful manifold of densities, potentials, and self-consistent solutions.

Despite its practical importance, this is often treated as an implementation detail rather than as a scientific problem.
Nonetheless, this starting point can affect both the robustness of self-consistent training and the transferability of the resulting functional.
While flexible neural representations are useful because they are expressive by construction, such flexibility also makes them harder to control.
The question of transferability is inherent to this problem, a network may reproduce a target accurately within the sampled training regime, but still fail when deployed on systems with different density distributions or bonding environments~\cite{dick_machine_2020}.
Imposing exact physical constraints, central to non-empirical functional design~\cite{pederson_difference_2023, beig_density_2003, sun_strongly_2015}, offers one way to control this behavior.
In machine-learned functionals, constraints can act not only as regularizers during training~\cite{pederson_difference_2023, nagai_machine-learning-based_2022, li_kohn-sham_2021}, but also as architectural priors that restrict the learned functional to a physically admissible self-consistent regime.
Although constrained neural architectures have shown clear benefits in differentiable-DFT settings~\cite{dick_highly_2021}, their role at the initialization stage remains insufficiently understood.

A natural way to study these questions is to consider the controlled problem of \emph{cloning} a known functional.
If a neural XC approximation is trained to reproduce an established functional, such as PBE, the target is known and errors can be evaluated systematically across systems, observables, numerical representations, and length scales.
This creates a testbed in which architectural choices can be isolated without the additional ambiguity of training against correlated reference data, where the exact functional is unknown and broad a posteriori validation is difficult.
Cloning is therefore useful not only as pre-training, but also as a diagnostic for whether a neural representation is physically admissible before it is used in self-consistent optimization.
It also provides a natural way to evaluate cross code portability.
As the learned functional remains anchored to a known parent approximation, its implementation follows the same local, grid-based evaluation paradigm used in standard XC libraries~\cite{lehtola_recent_2018, soler_libgridxc_nodate}.

In this work, we use functional cloning to address three questions relevant to machine-learned physical models embedded in self-consistent equations.
(i) Does enforcing exact constraints improve a neural clone even when the target functional already satisfies those constraints?
(ii) How does the numerical realization of the density affect the portability of a learned functional across all-electron and pseudopotential codes?
(iii) Do cloned and constrained models provide better starting points for subsequent self-consistent optimization and better transfer from molecules to crystalline solids?
The remainder of the paper is organized accordingly.
Section~\ref{sec:cloning} defines the cloning framework, neural GGA representation, descriptor transformations, constraints, and molecular sampling protocol.
Section~\ref{sec:cloning_results} presents the cloning results and cross-code density-representation analysis.
Section~\ref{sec:full_training} examines self-consistent optimization from cloned and random initializations, including molecular-to-solid transfer.
Section~\ref{sec:conclusions} summarizes the implications for stable and transferable machine-learned XC design.

\section{\label{sec:cloning} Cloning exchange--correlation functionals as a controlled testbed}

\subsection{\label{ssec:cloning_framework} Cloning framework}

Our initial assumption is that, in order to optimize an XC functional, and particularly if we need to limit the number of SCF cycles during optimization to reduce computational cost, it is better to start from an already self-consistent functional.
Therefore, we propose a strategy where we start from a controlled initialization problem in which a NN-XC functional is trained to reproduce a known target.
Related supervised-learning strategies have previously been used to reproduce existing exchange--correlation approximations\cite{ryabov_application_2022}, for example, in the de-orbitalization of SCAN, where a neural network was trained to replicate a meta-GGA functional from density and local derivative information without explicit dependence on the orbital kinetic-energy density~\cite{pokharel_exact_2022}.
Here, however, we use cloning not primarily to build a surrogate for a higher-rung functional, but to define a controlled diagnostic and initialization framework for NN-XC functionals.
At this stage the objective is not to improve upon the reference functional, but to assess how specific choices for training---including the choice of input parameters and enforcement of known physical constraints---affect the ability of the NN-XC to recover a stable, well-defined functional.
Because the resulting model can be used as the starting point for subsequent
XC optimization, this cloning step also plays the
role of a pre-training stage in our broader workflow.

In this work we focus on semilocal GGA-level functionals, for which the exchange--correlation energy density can be written in terms of enhancement factors multiplying the corresponding uniform-electron-gas reference energy densities ($\epsilon^{\mathrm{UEG}}$).
This representation provides a particularly controlled setting for functional cloning: the neural network is not asked to learn an arbitrary total energy, but rather a local multiplicative correction with a known analytical target and well-defined limiting behavior.
Here we restrict our study to GGA-level cloning.
However, this framework also naturally extends to other functionals by augmenting the descriptor set, for example with kinetic-energy-density-dependent variables in the meta-GGA case.
For hybrid functionals, the semilocal exchange--correlation component could be cloned in the same way, while the nonlocal exact-exchange contribution would remain a separate fixed contribution.
Within this GGA-level setting, the cloned models are constructed from local and semilocal ingredients evaluated on the numerical integration grid.
Specifically, they receive as inputs the total density $\rho(\mathbf r)$, the spin polarization
\[
\zeta(\mathbf r) =
\frac{\rho_\uparrow(\mathbf r)-\rho_\downarrow(\mathbf r)}
{\rho(\mathbf r)},
\]
where $\rho_\uparrow$ and $\rho_\downarrow$ are the spin-up and spin-down densities, and the density gradient
$\nabla \rho(\mathbf r) = (\partial_x \rho, \partial_y \rho, \partial_z \rho)$.

\clearpage   % flush all floats
\begin{widetext}
\begin{figure*}[t]
\centering
\includegraphics[width=\linewidth]{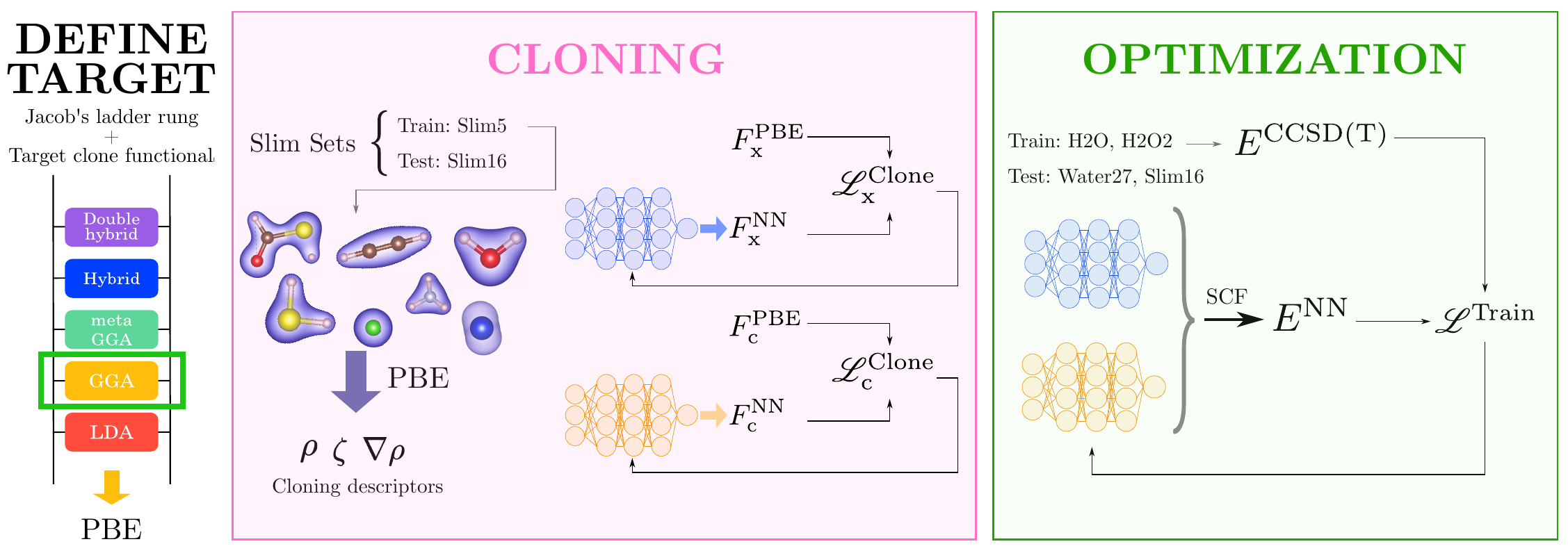}
\caption{
Scheme of the complete workflow for the construction and optimization of the machine-learned exchange--correlation functional.
It includes three stages: choosing the functional form and target rung of Jacob's ladder, cloning a known functional, and using the cloned model as an initialization for self-consistent optimization.
In this work, we choose a GGA-level representation and PBE as the cloning target.
The cloning stage consists of training a network to match the enhancement factor of the chosen reference functional (PBE).
This is done by fitting two different neural networks ---one for exchange (blue) and one for correlation (yellow)---, in selected grid points (cloning descriptors).
These last are obtained from ground state densities, obtained with the chosen functional, in the train set molecular systems.
In the subsequent self-consistent optimization stage, both networks are optimized jointly, starting from the cloned parameters.
In contrast to cloning, each optimization step requires a full self-consistent-field (SCF) cycle with the neural functional, making the model part of the nonlinear fixed-point problem.
}
\label{fig:scheme_pt_training}
\end{figure*}
\end{widetext}

Defining the local GGA descriptor vector as
\begin{equation}
\mathbf{x}(\mathbf r)
=
\left\{
\rho(\mathbf r),
\zeta(\mathbf r),
\nabla \rho(\mathbf r)
\right\},
\end{equation}
the GGA exchange--correlation energy density can be written as
\begin{align}
\epsilon_{\mathrm{xc}}^{\mathrm{GGA}}(\mathbf r)
&=
\epsilon_{\mathrm{x}}^{\mathrm{GGA}}(\mathbf r)
+
\epsilon_{\mathrm{c}}^{\mathrm{GGA}}(\mathbf r)
\nonumber \\
&=
F_{\mathrm{x}}\!\left(\mathbf{x}(\mathbf r)\right)
\epsilon_{\mathrm{x}}^{\mathrm{UEG}}[\rho,\zeta](\mathbf r)
+
F_{\mathrm{c}}\!\left(\mathbf{x}(\mathbf r)\right)
\epsilon_{\mathrm{c}}^{\mathrm{UEG}}[\rho,\zeta](\mathbf r).
\end{align}
The cloning task is just a pointwise supervised-learning problem in which the neural networks are trained to reproduce known exchange and correlation enhancement factors, $F_{\mathrm{x}}$ and $F_{\mathrm{c}}$.
Importantly, we fit the exchange and correlation components separately.
This decomposition allows us to assess the errors associated with each
enhancement factor independently avoiding possible error cancellations. 
Hence, our NN-XC functional is represented by two neural networks: one for
exchange and one for correlation.
We choose PBE~\cite{perdew_generalized_1996,burke_perspective_2012} as the
target functional.
Its functional form was derived by enforcing exact physical constraints on the exchange--correlation energy.
Its enhancement factors therefore provide a natural target for assessing whether a limited-size neural network can reproduce a physically constrained semilocal enhancement factor, along with its self-consistent energy and density.
The cloning is just a training where one minimizes the point-wise discrepancy between the neural network output and PBE enhancement factors at any given combination of local density variables.
Accordingly, the training loss is defined as
\begin{equation}
    \mathscr{L}^{\text{Clone}}_{\text{x/c}} = 
    \sum_i 
    \left( F^{\text{NN}}_{\text{x/c}}(\rho_i, \zeta_i, \nabla \rho_i) -
    F_{\text{x/c}}^{\text{PBE}}(\rho_i, \zeta_i, \nabla \rho_i)\right)^2,
\end{equation}
where the index $i$ labels the descriptor points
$(\rho_i,\zeta_i,\nabla\rho_i)$ used for cloning.
These descriptor points can be taken from real-space grids obtained from ground state densities, or generated synthetically in order to sample the
descriptor space in a controlled way, as the loss depends only on local descriptor values and does not explicitly involve the spatial coordinate $\mathbf r_i$.
This makes the cloning procedure remarkably flexible and turns it into a controlled way to test how descriptor sampling, architectural constraints, and the separation of exchange and correlation affect accuracy and transferability.

We summarize the cloning framework within in Fig.~\ref{fig:scheme_pt_training}.
The figure shows the complete workflow, beginning with the choice of a rung of
Jacob's ladder and a target functional, followed by the cloning of the exchange
and correlation enhancement factors using two separate neural networks.
The final optimization stage, which is described in
Sec.~\ref{sec:full_training}, is also shown in the figure for context.
This last stage consists of further training our networks against accurate reference data, using the cloned networks as the starting point.

\subsection{\label{ssec:cloning_model} Neural functional form, descriptor transformations, and constraints}
The construction of a clone model follows the strategy introduced in our previous work~\cite{dick_highly_2021}, adapted here to the pointwise reproduction of a known semilocal functional.
As discussed in Sec.~\ref{ssec:cloning_framework}, the models considered here reproduce a functional at the GGA level.
Their inputs are therefore restricted to the density $\rho(\mathbf r)$, the spin polarization $\zeta(\mathbf r)$, and the density gradient $\nabla\rho(\mathbf r)$, with spin-resolved components included in the spin-polarized case.
To improve numerical stability and learning efficiency, we first transform these raw physical inputs into variables with more suitable numerical ranges.

In particular, we introduce the descriptors
\begin{align}
  x_0= & \rho^{1/3}, \\
   x_1 = & \frac{1}{2}\left\{ (1+\zeta)^{4/3}+ (1-\zeta)^{4/3}\right\}, \\
  x_2 = &s = \frac{|\nabla \rho|}{2k_F\rho},
\end{align}
where $k_F=(3\pi^2\rho)^{1/3}$ is the Fermi momentum.
In addition, we further compress the dynamic range of the descriptors by applying logarithmic transformations, 
\begin{align}
    \tilde{x}_0= &\log \left(x_0+\varepsilon_{\log }\right), \\
    \tilde{x}_1=& \log \left(x_1+\varepsilon_{\log }\right), \\
    \tilde{x}_2=&\left\{1-e^{-(x_2)^2}\right\} \log \left(x_2+1\right),
\end{align}
where $\varepsilon_{\log} = 10^{-5}$ is introduced to ensure numerical stability. 
The specific transformation chosen for $\tilde{x}_2$, following Ref.~\cite{dick_highly_2021}, guarantees that its derivative vanishes at $x_2 = 0$. 
This induces a soft constraint on the enhancement factors $F_{\mathrm{x/c}}$, promoting the correct behavior in the slowly varying density limit. 
Rather than enforcing physical constraints through penalty terms in the loss function~\cite{pokharel_exact_2022}, we incorporate them directly into the functional form of the model. 
To do so, we map the raw neural-network outputs $\mathcal{F}_{\mathrm{x/c}}$ through the following transformations so that selected exact conditions are satisfied by construction,
\begin{subequations}\label{eq:ph_constraints}
\begin{align}
F_{\mathrm{x}}(\tilde{x}_2)
&= 1 + I_{1.804}\!\left[\tilde{x}_2\,
\mathcal{F}_{\mathrm{x}}\left(\tilde{x}_2;\boldsymbol{\omega}_{\mathrm{x}}\right)\right],
\label{eq:ph_constraints_x} \\
F_{\mathrm{c}}(\tilde{x}_0,\tilde{x}_1,\tilde{x}_2)
&= 1 + I_2\!\left[\tilde{x}_2\,
\mathcal{F}_{\mathrm{c}}\left(\tilde{x}_0,\tilde{x}_1,\tilde{x}_2;\boldsymbol{\omega}_{\mathrm{c}}\right)\right].
\label{eq:ph_constraints_c}
\end{align}
\end{subequations}
where $I_a(x)$ maps its argument to the finite interval $[-1, a-1]$, 
\begin{equation}
I_a(x)=\frac{a}{1+(a-1) \exp (-x)}-1
\end{equation}
while preserving $I_a(0)=0$. 
This guarantees that, at $s=0$, the enhancement factors satisfy $F_{\mathrm{x}}=F_{\mathrm{c}}=1$, thereby recovering the uniform-electron-gas limit.
This construction automatically enforces the Lieb--Oxford bound for exchange, $0 \le F_{\mathrm{x}} \le 1.804$~\cite{lieb_improved_1981,perdew_gedanken_2014}
%, as well as the non-negativity of $F_{\mathrm{c}}$.
%
Recently, the bound has been lowered to 1.5765~\cite{lewin_improved_2022}, but since our goal is to clone PBE, we keep the same bound used in the original work~\cite{perdew_generalized_1996}.
The upper bound on the correlation enhancement factor is not physically required; it is imposed purely as a regularization to prevent unphysical extrapolations during training.
Thereby, it should not be interpreted as a fundamental exact bound analogous to the Lieb--Oxford constraint on exchange.
Finally, making $F_{\mathrm{x}}$ depend only on $\tilde{x}_2$ ensures that the exchange functional obeys the spin-scaling relation and the uniform density-scaling behavior of $E_{\mathrm{x}}$.
We implement this functionality in our package \texttt{xcquinox-clone}~\cite{navarro-rodriguez_xcquinox-clone_nodate}, based on \texttt{xcquinox}~\cite{wills_xcquinox_nodate}, which builds on \texttt{equinox} and \texttt{JAX} and includes the neural-network architecture, descriptor transformations, constrained output mappings, and interfaces to the DFT package \textsc{PySCF} and its differentiable extension \textsc{PySCFAD}~\cite{zhang_differentiable_2022}.
The exchange and correlation enhancement factors are modeled using independent
multilayer perceptrons implemented with \texttt{equinox} and using an identical number of hidden layers and neurons-per-layer.
This choice keeps the model capacity fixed when comparing exchange and
correlation errors, so that differences in performance can be attributed to the
learned enhancement factors rather than to different network sizes.
Nevertheless, the implementation of the two models is not identical: the exchange network only depends on $\tilde{x}_2$, so it receives a single input, whereas the correlation network receives the full descriptor set $(\tilde{x}_0,\tilde{x}_1,\tilde{x}_2)$ (three inputs).

We use compact networks consisting of three hidden layers with 16 neurons per layer and GELU activation functions, and keep this architecture fixed throughout both the cloning and full-training stages.
This allows us to compare different initialization strategies, constraint choices, and training datasets
without confounding these effects with changes in model capacity.
The networks are cloned using the Adam optimizer with a learning-rate schedule consisting of an initial constant phase at $10^{-3}$, followed by a linear decay to $10^{-5}$ between 50\% and 90\% of the total number of optimization steps, and a final constant phase. 
Training is performed for $2\times10^4$ epochs to ensure convergence.
Throughout the paper, we use the following nomenclature for the neural functionals.
Cloned PBE models are denoted PBE-Cl-C and PBE-Cl-U, where C and U stand for constrained and unconstrained architectures, respectively.
When the density source must be specified, we append it explicitly as PBE-Cl-C/Py or PBE-Cl-C/SIE, indicating descriptors generated with \textsc{PySCF} or \textsc{siesta}.
When no density source is shown, \textsc{PySCF} descriptors are implied.
Models that are subsequently self-consistently optimized against correlated reference energies are denoted XCQ-C or XCQ-U, with the initialization indicated in brackets: [Cl] for cloned initialization and [Rd] for random initialization.
Random networks evaluated without cloning or self-consistent optimization are denoted Rd-C or Rd-U.

\subsection{\label{ssec:cloning_data} Molecular datasets and descriptor sampling protocol}

We use the Slim benchmark sets as train and test sets~\cite{gould_slim_2025}.
These are down-selections of the full GMTKN55\cite{goerigk_look_2017} benchmark set with a maximum number of atoms within each subset.
The descriptors $(\rho,\zeta,\nabla\rho)$ are obtained from PBE ground state densities, in order to have physically meaningful combinations of the input quantities.
Although the Slim sets are organized in terms of energy differences, the cloning loss is defined pointwise in descriptor space.
We therefore treat each molecular geometry as an independent source of density points and remove repeated molecules within each subset to minimize duplicate entries (see Supplementary Sec.~S1 for more details).
For cloning, a subset of $N_m$ molecular systems is randomly selected from Slim05, from which a total of $N_p$ points are sampled for the cloning procedure. 
For each molecule, $N_p/N_m$ points are drawn from the real-space grid. 
The sampling is performed with probability proportional to $w_i \rho_i$, where $w_i$ are the integration weights associated with each grid point $\mathbf{r}_i$ in \textsc{PySCF}. 
This choice follows from the discrete form of the XC energy,
\begin{equation}
    E_{\mathrm{xc}} = \sum_i w_i \rho_i \epsilon_{\mathrm{xc},i},
\end{equation}
and biases the descriptor sampling toward regions that contribute most strongly to the energy.
This weighting is particularly useful for atom-centered quadrature grids, where the integration points are not uniformly distributed in space.
It also gives more weight to the points that contribute most to the integrated XC energy.
The cross-code comparison presented in the next section provides an explicit test of how this density-weighted sampling interacts with the difference between all-electron and pseudopotential density representations.

As we follow a random selection of molecules and grid points, we can generate cloned models with statistical variability.
The presented results are obtained by repeating the cloning procedure $R$ times using independent descriptor samples and reporting averaged metrics over these repetitions.
We use the Slim16 set for evaluation and testing.
This provides a more demanding test than the training set because Slim16 contains molecules with up to sixteen atoms, whereas Slim05 contains molecules with at most five atoms.
It also reduces the effective overlap between training and test molecular systems.
Although the two sets are not guaranteed to be strictly disjoint, the random selection of molecular systems used for cloning and the large number of entries in each set make any residual overlap statistically negligible and unlikely to bias the results.
All \textsc{PySCF} calculations used to generate reference densities and descriptor values are performed with the def2-TZVP basis set, a grid level of 4, and a convergence tolerance of $10^{-9}$ Hartree.
We should note that, in the cloning stage, \textsc{PySCF} is used only to obtain the fixed descriptor points and the corresponding reference enhancement factors.
It is not part of the training loop: the loss is evaluated directly on the precomputed descriptor samples and therefore does not require self-consistent field calculations with the neural functional.

\subsection{Cross-code deployment in \textsc{siesta}}
\label{ssec:siesta_implementation}

To assess whether a NN-XC functional trained in one electronic-structure environment can be deployed in another, we implemented the same neural functional in \textsc{siesta}.
This provides a stringent portability test because the densities used by \textsc{PySCF} and \textsc{siesta} are represented in fundamentally different ways: \textsc{PySCF} uses all-electron Gaussian basis sets and atom-centered quadrature grids, whereas \textsc{siesta} uses norm-conserving pseudopotentials, numerical atomic orbitals, and real-space uniform grids.
In the portability tests reported below, the network architecture and weights are kept fixed; only the electronic-structure code used to evaluate the density, descriptors, and self-consistent potential is changed.

The \textsc{siesta} implementation follows the standard structure of a GGA XC functional, as it is done in libGridXc\cite{soler_libgridxc_nodate}.
At each grid point, the code evaluates the transformed descriptors, the exchange and correlation neural-network outputs, and the corresponding enhancement factors.
Because self-consistent calculations require not only the XC energy density but also the derivatives entering the potential, the derivatives of the descriptor transformations, output mappings, and neural-network operations were implemented explicitly.
Additional implementation details and validation tests are provided in the Supplementary Information.

\section{\label{sec:cloning_results} Cloning results}
\subsection{\label{ssec:results_constraints_clone} Role of constraints}

We first assess how the architectural constraints introduced in
Sec.~\ref{ssec:cloning_model} affect the ability of the neural functional to reproduce the reference PBE functional.
We compare two otherwise identical cloned NN-XC models. (i) the constrained clone PBE-Cl-C, in which
the output transformations of Eqs.~\ref{eq:ph_constraints_x} and
\ref{eq:ph_constraints_c} are applied, and (ii) the unconstrained clone PBE-Cl-U, in which these constraints are not imposed.
The goal is to evaluate whether enforcing known physical conditions leads to a more faithful and stable reproduction of the
reference functional.

The amount of descriptor data used for cloning was selected from the convergence analysis reported in Supplementary Sec.~S2.
In that analysis, we monitored the mean absolute per-electron energy difference
with respect to PBE as a function of the number of molecular systems, $N_m$, and the
total number of sampled grid points, $N_p$.
The error decreased as both quantities were increased and saturated for the
settings used here, $N_m=25$ and $N_p=20{,}000$.
These values were therefore kept fixed for all cloning runs.

Because the random selection of molecules and grid points introduces sampling
variability, we repeated the cloning procedure five times for each architecture.
Each repetition used a different random subset of $N_m=25$ molecular systems from
Slim05, listed in Supplementary Table~S1.
For each repetition, the constrained and unconstrained models were trained on exactly the same descriptor points, so that the only difference between the two
sets of models were the presence or absence of the constraint-enforcing output
mapping.
The cloning losses and validation scatter plots are reported in the Supplementary
Figs.~S2 and S3, confirming that both architectures were trained to stable solutions.
Furthermore, the ability of the networks to reproduce PBE is confirmed in Supplementary Sec.~S6.

We evaluate the cloned functionals using two complementary metrics.
The first is the mean absolute total-energy error with respect to PBE,
\begin{equation}
    \mathrm{MAE}_{\mathrm{TE}}
    =
    \frac{1}{M}
    \sum_{i=1}^{M}
    \left| E^{\mathrm{NN}}_i - E^{\mathrm{PBE}}_i \right|,
\end{equation}
where $M$ is the number of molecular systems in the Slim16 test set.
Although absolute total-energy errors are not a standard metric for comparing
independent XC approximations, they are appropriate here
because the objective is to reproduce PBE in the same computational setting.
A successful clone should therefore recover not only the pointwise enhancement
factors, but also the self-consistent density and total energy obtained with the reference functional.
As shown in Supplementary Sec.~S7, density-driven errors are small in this
setting, so $\mathrm{MAE}_{\mathrm{TE}}$ primarily reflects the integrated effect of residual functional differences over the molecular densities.

The second metric is WTMAD-2~\cite{goerigk_look_2017}, evaluated here with PBE
as the reference rather than high-level wavefunction data,
\begin{equation}
\mathrm{WTMAD\mbox{-}2}
=
\frac{\overline{|\Delta E_{\mathrm{PBE}}|}_{\mathrm{mean}}}
{N_{\mathrm{total}}}
\sum_{i=1}^{N_{\mathrm{subsets}}}
\frac{N_i\,\mathrm{MAD}_i}
{\overline{|\Delta E_{\mathrm{PBE}}|}_i},
\end{equation}
where $i$ indexes the Slim16 subsets, $N_i$ is the number of entries in subset
$i$, and $N_{\mathrm{total}}=\sum_i N_i$.
Here,
\begin{equation}
\mathrm{MAD}_i
=
\frac{1}{N_i}
\sum_{j=1}^{N_i}
\left|
\Delta E^{\mathrm{NN}}_{ij}
-
\Delta E^{\mathrm{PBE}}_{ij}
\right|,
\end{equation}
and $\overline{|\Delta E_{\mathrm{PBE}}|}_i$ is the mean absolute PBE energy
difference in subset $i$, while
$\overline{|\Delta E_{\mathrm{PBE}}|}_{\mathrm{mean}}$ is the corresponding
mean over all Slim16 entries.
This metric measures how closely the cloned functionals reproduce PBE energy differences across chemically diverse subsets.

\begin{figure}[h]
    \centering
\includegraphics[width=\linewidth]{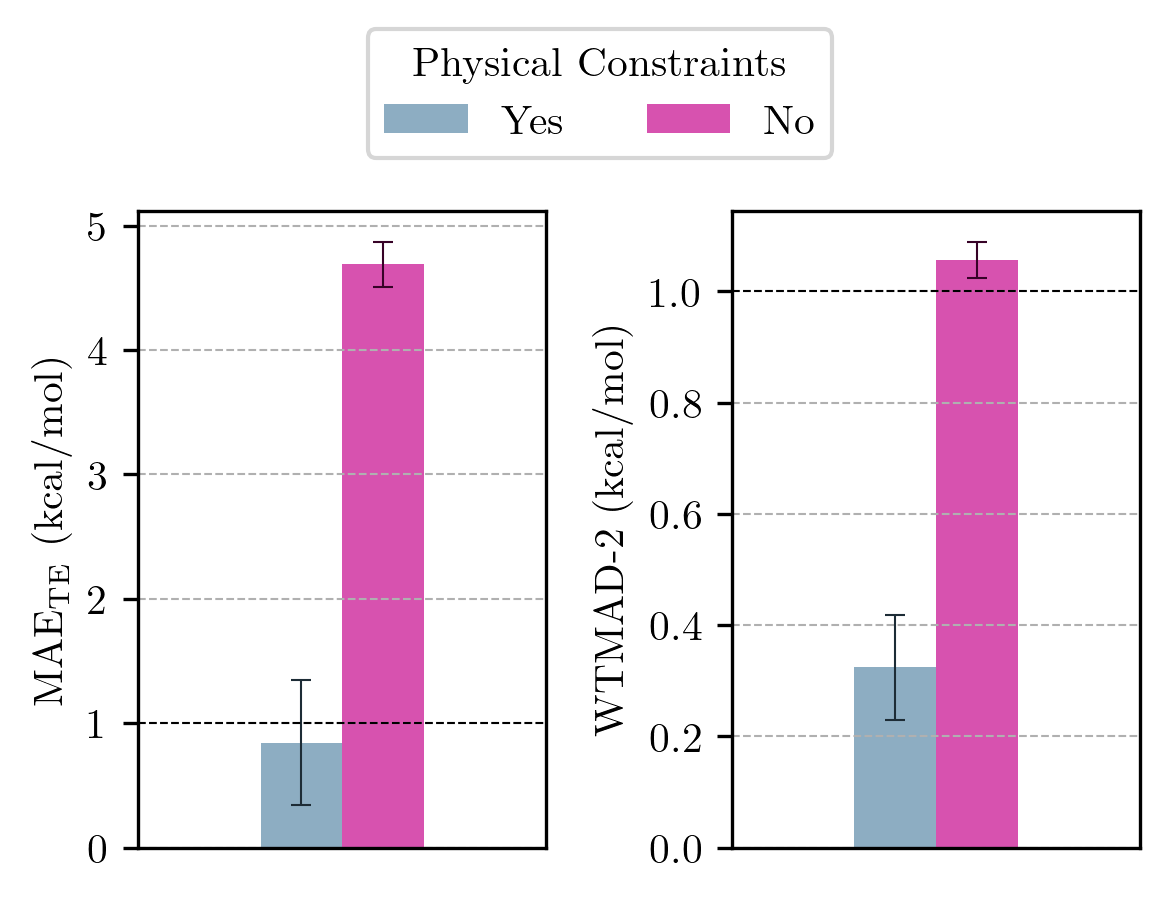}
\caption{Evaluation of self-consistent energies obtained with the constrained
(blue) and unconstrained (fuchsia) cloned networks with respect to PBE.
We report the mean absolute total-energy error ($\mathrm{MAE}_{\mathrm{TE}}$)
and WTMAD-2.
Averages are taken over all entries in Slim16 and over the five independent
cloning runs; error bars indicate the standard deviation across these five
runs.
The dashed black line marks 1 kcal/mol as a visual reference.}
\label{fig:tl_vs_unct}
\end{figure}

Fig.~\ref{fig:tl_vs_unct} compares the constrained and unconstrained clones
using the two metrics defined above.
The complete numerical values, including the results for each independent
cloning run, are given in Supplementary Tables~S2 and S3.
Both metrics show that enforcing physical constraints improves the fidelity
of the cloned functional.
The effect is particularly pronounced for $\mathrm{MAE}_{\mathrm{TE}}$, where the error decreases from nearly 5 kcal/mol for the unconstrained network to
approximately 0.8 kcal/mol for the constrained network.
The WTMAD-2 errors are smaller in magnitude, as expected for energy-difference
metrics where systematic contributions to total energies partially cancel, but the constrained architecture again gives the lower error.

Overall, these results show that the constraint-enforcing architecture improves
PBE cloning, even though the reference functional itself already satisfies the
same exact conditions.
This indicates that constraints restrict the neural representation in a way that yields a more accurate and self-consistent reproduction of the target functional.

\subsection{\label{ssec:portability_clone} Dependence on the numerical realization of the density}
% PySCF vs SIESTA

The previous subsection assessed the role of constraints when the descriptors
used for cloning and the self-consistent evaluation were generated within the
same electronic-structure environment.
Here we ask a different question: how sensitive is the cloned functional to the
numerical realization of the density used to construct its descriptor space?
This issue is important because a density is not represented uniquely in a
practical calculation.
It depends on choices such as the basis set, integration grid, self-consistent
convergence thresholds, and, most importantly for the present comparison, the
use of an all-electron or pseudopotential description.

We therefore compare cloning and evaluation across \textsc{PySCF} and
\textsc{siesta}.
uses norm-conserving pseudopotentials, numerical atomic orbitals, and real-space uniform
grids.
Even when the basis quality and integration settings are chosen to be broadly
comparable, the all-electron and pseudopotential densities are intrinsically
different, especially in the core region.
As a result, a NN-XC functional cloned on descriptors generated in one code may
not evaluate the same functional when deployed in another.

To test this effect, we trained a constrained network using descriptors obtained
from PBE ground-state densities generated with \textsc{siesta}; we refer to this
model as PBE-Cl-C/SIE.
We compare it with the constrained model discussed in
Sec.~\ref{ssec:results_constraints_clone}, which was cloned using
\textsc{PySCF} descriptors and is denoted PBE-Cl-C/Py.
Both models use the same architecture and were cloned using the same molecular
geometries, with $N_m=25$ and $N_p=20{,}000$.
Because the purpose of this comparison is diagnostic rather than statistical, we performed a single cloning run for the PBE-Cl-C/SIE model.

Each network was then evaluated self-consistently in both codes on the Slim16
test set~\cite{gould_slim_2025}.
In each case, the NN-XC results were compared with the PBE results obtained in
the same code.
This protocol separates two effects: the dependence of the clone on the density
representation used during training, and the portability of the trained network
when evaluated in a different electronic-structure code.

\begin{figure}[h!]
    \centering
\includegraphics[width=\linewidth]{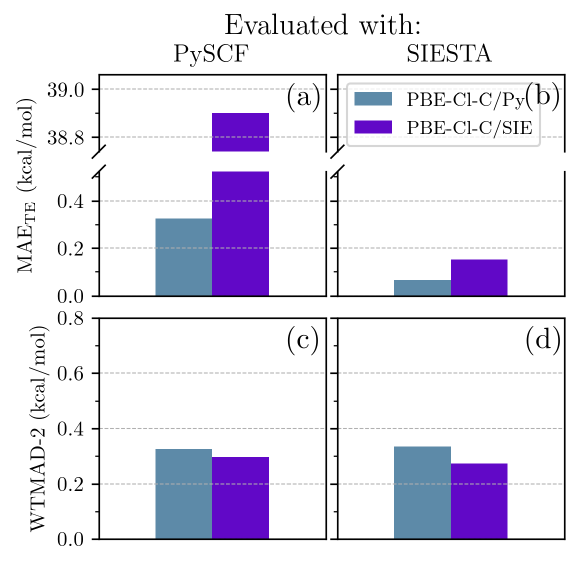}
    \caption{Mean absolute total-energy difference ($\mathrm{MAE}_{\mathrm{TE}}$)
and WTMAD-2 for the NN-XC functionals compared with PBE, evaluated in
\textsc{PySCF} (left, panels a and c) and in \textsc{siesta} (right, panels b
and d).
Blue bars correspond to constrained NNs cloned using densities from
\textsc{PySCF}; purple bars correspond to constrained NNs cloned using densities
from \textsc{siesta}.
All errors are evaluated on Slim16, using PBE from the corresponding code as the
reference.}
    \label{fig:ev_SIESTA_PYSCF}
\end{figure}

The results are summarized in Fig.~\ref{fig:ev_SIESTA_PYSCF}.
For WTMAD-2, the two cloned models give comparable errors in both codes,
indicating that energy differences are relatively insensitive to the
code used to generate the cloning descriptors.
This is consistent with the fact that core contributions largely cancel in
energy differences, while the valence-density regions that dominate chemical
energy changes are represented in both all-electron and pseudopotential
calculations.

The total-energy metric shows a more asymmetric behavior.
The PBE-Cl-C/Py model gives small $\mathrm{MAE}_{\mathrm{TE}}$ values when
evaluated in either \textsc{PySCF} or \textsc{siesta}, and the
PBE-Cl-C/SIE model also performs well when evaluated in \textsc{siesta}.
In contrast, the PBE-Cl-C/SIE model evaluated in \textsc{PySCF} gives a
much larger total-energy error, close to 40 kcal/mol.
This behavior is expected: a model cloned on pseudopotential densities has never
seen the all-electron core-density regime and therefore cannot be expected to
reproduce all-electron total energies accurately.
By contrast, when evaluated in a pseudopotential code, the all-electron-trained model is applied only to the valence-like density regime.

These results show that the numerical realization of the density matters for
functional cloning.
For total-energy reproduction, portability is asymmetric: an all-electron clone
can transfer to a pseudopotential setting more reliably than a pseudopotential clone can transfer back to an all-electron calculation.
For energy differences, however, the dependence is much weaker, because the missing or altered core-density contributions largely cancel.
This distinction is important for interpreting cloned NN-XC functionals across
codes: apparent portability depends not only on the functional form and weights,
but also on which parts of the density space are sampled by the training and
deployment environments.

\section{\label{sec:full_training} Self-consistent optimization from cloned initializations}

\subsection{\label{ssec:full_training_framework} Optimization framework}
% CCSD(T), SCF in the loop, loss over systems

Here, we use the cloned functionals as
initializations for self-consistent optimization against higher-level reference
data.
Although both cloning and optimization update the parameters of the same
neural functional, they differ in both objective and computational structure.
During cloning, the network is trained to reproduce a known functional
locally, at fixed descriptor points, and the optimization is therefore decoupled from the self-consistent electronic-structure problem.
While in the optimization stage the target is a ground-state observable obtained after solving SCF Kohn--Sham
equations with the neural functional that is being optimized.

In this proof-of-concept study, our reference data are molecular total energies
computed at the CCSD(T) level (coupled cluster with singles, doubles, and perturbative triples).
For a set of training systems indexed by $k$, we define the self-consistent
training loss as
\begin{equation}
\mathscr{L}^{\mathrm{Train}}(\boldsymbol{\theta})
=
\sum_{k}
\left[
E_k^{\mathrm{NN}}(\boldsymbol{\theta})
-
E_k^{\mathrm{ref}}
\right]^2,
\label{eq:full_training_loss}
\end{equation}
where $\boldsymbol{\theta}$ denotes the combined parameters of the exchange and
correlation networks.
Here, $E_k^{\mathrm{NN}}(\boldsymbol{\theta})$ is the total energy obtained from
a SCF calculation with the neural XC functional, while
$E_k^{\mathrm{ref}}$ is the corresponding CCSD(T) reference energy.
Unlike in the cloning stage, the loss depends on the neural-network
parameters both explicitly through the XC energy density and implicitly through
the self-consistent density.

During the optimization, the exchange and correlation networks are updated
jointly.
Their architectures and constraint-enforcing output transformations are kept
fixed; only the network parameters are changed.
This allows us to test how much the functional can move away from its cloned PBE
initialization when trained on a small  dataset, and how this behavior
depends on the quality and physical constraints of the initial clone.
The relation between the cloning and self-consistent optimization stages is
shown schematically in Fig.~\ref{fig:scheme_pt_training}.

We use \textsc{PySCFAD}~\cite{zhang_differentiable_2022} as our DFT engine for the optimization.
This is because it provides automatic differentiation through the self-consistent procedure, and it is easy to integrate in our own \texttt{xcquinox}~\cite{wills_xcquinox_nodate} framework.
At each optimization step, a complete SCF calculation is performed with the
current neural functional, the loss in Eq.~\ref{eq:full_training_loss} is
evaluated, and gradients with respect to the neural-network parameters are used
to update the model.
This makes the self-consistent optimization much more computationally demanding than cloning, and we restrict this stage to 300 optimization epochs.
We use the Adam optimizer with a piecewise learning-rate schedule: the learning
rate is initially $10^{-3}$, remains constant until epoch 30, then decays
linearly to $10^{-6}$ by epoch 200, and is kept fixed at that value for the
remainder of the optimization.

\subsection{\label{ssec:init_effect} Effect of initialization and constraints on self-consistent optimization}
% constrained clone vs unconstrained clone vs random

We have shown that architectural constraints improve the fidelity of PBE cloning.
We now ask whether the same ingredients also help once the neural functional is optimized against correlated reference data.
In particular, we address two related questions: whether starting from a cloned, self-consistent functional improves the stability and transferability of the subsequent optimization, and whether imposing physical constraints remains beneficial after the model is allowed to move away from PBE.

To test this, we optimize four NN-XC functionals that differ in both initialization and architecture.
The initial parameters are either taken from the corresponding PBE-cloned model or generated randomly.
For each initialization, we consider both constrained and unconstrained architectures.
The four optimized models are denoted XCQ-C[Cl], XCQ-U[Cl], XCQ-C[Rd], and XCQ-U[Rd], where the bracket identifies whether the optimization starts from a PBE clone or from random parameters.
All models are optimized against CCSD(T) reference energies computed with \textsc{PySCF}, using the same def2-TZVP basis set and $10^{-9}$ Hartree convergence tolerance used for the PBE calculations.
Because this stage is intended as a small proof of concept, the training set is deliberately minimal: it contains $\mathrm{H}_2\mathrm{O}$, $\mathrm{H}_2\mathrm{O}_2$, and the corresponding reaction energy.
The training losses for the four models are shown in Supplementary Fig.~S6.

We then evaluate how this minimal self-consistent optimization transfers beyond the training systems.
As a near-domain test, we use the WATER27 subset of GMTKN55~\cite{bryantsev_evaluation_2009,anacker_new_2014}, which contains binding energies of neutral, protonated, and deprotonated water clusters.
As a more demanding out-of-domain test, we use the non-spin-polarized molecular systems of Slim16, which contain chemical environments and atomic species not present in the training set.
We restrict the Slim16 evaluation to unpolarized molecules because both training systems have $\zeta=0$ everywhere, and the optimization therefore introduces no new information about spin-polarized densities.
For both test sets, we report the mean absolute total-energy error, $\mathrm{MAE}_{\mathrm{TE}}$, and the mean absolute deviation of the relevant energy differences, MAD, using CCSD(T) as the reference.

\begin{figure}[h]
    \centering
    \includegraphics[width=\linewidth]{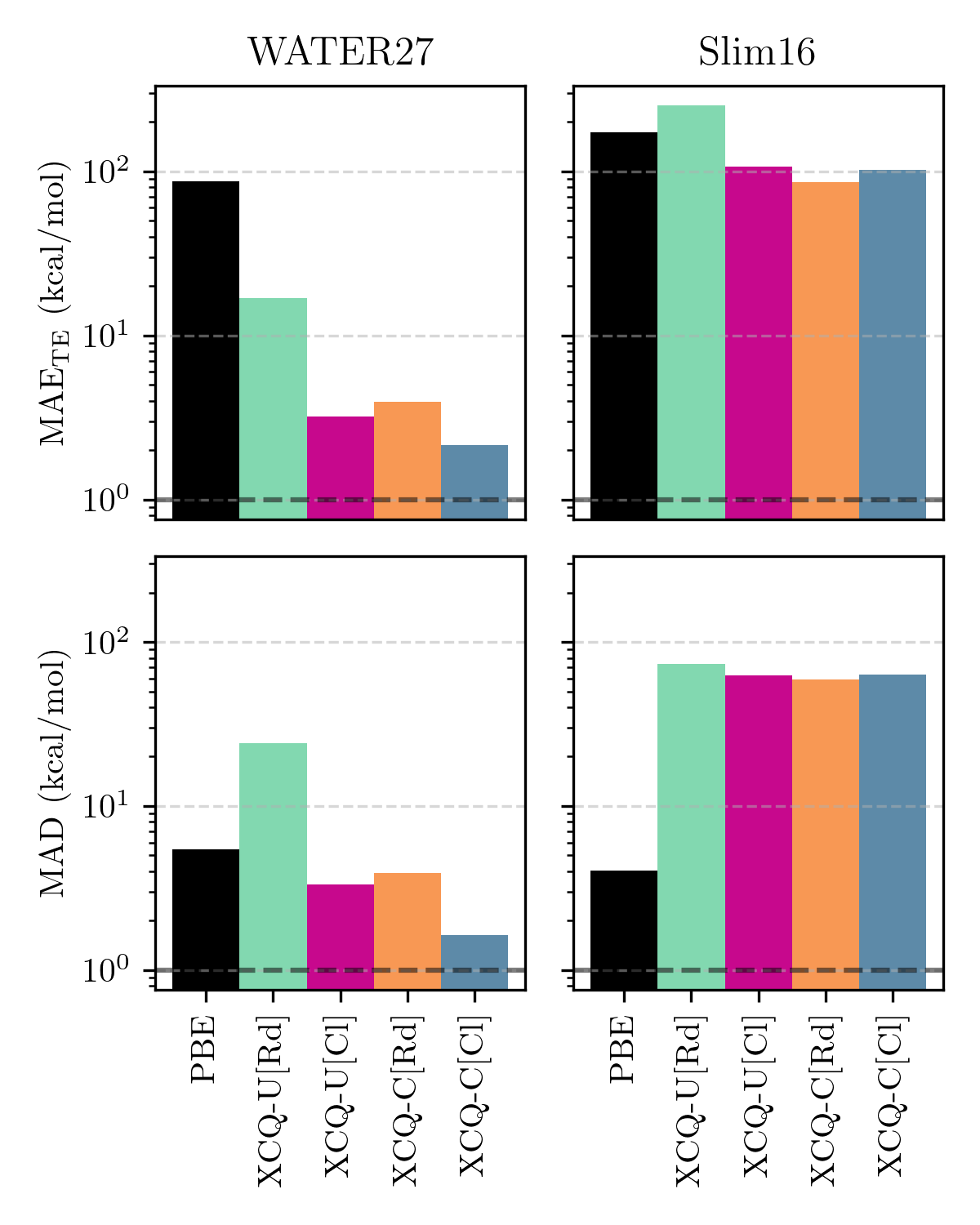}
    \caption{Evaluation of self-consistently optimized NN-XC functionals, denoted XCQ.
    We compare XCQ-C[Cl], XCQ-U[Cl], XCQ-C[Rd], and XCQ-U[Rd], corresponding to constrained (C) and unconstrained (U) architectures initialized either from cloned PBE models or from random parameters.
    The models are evaluated using the mean absolute total-energy error ($\mathrm{MAE}_{\mathrm{TE}}$) and the mean absolute deviation of energy differences (MAD) with respect to CCSD(T) reference energies for WATER27 and for the non-spin-polarized molecular systems of Slim16.
    PBE is included as a reference.}
    \label{fig:full_training}
\end{figure}

The results are shown in Fig.~\ref{fig:full_training}; the corresponding individual quantities are reported in Supplementary Figs.~S9 and S10.
For WATER27, all optimized networks improve over PBE in total energies.
For the energy-difference metric, all models except the unconstrained network initialized from random parameters improve over PBE.
This failure is consistent with the uneven behavior of that model during training: although it reaches a small error for $\mathrm{H}_2\mathrm{O}_2$, the error for $\mathrm{H}_2\mathrm{O}$ remains roughly two orders of magnitude larger.
The result illustrates the difficulty of self-consistently optimizing an unconstrained neural functional from random parameters using such a small training set.

The clearest trend in the near-domain test is that cloned initializations improve the final optimized models.
For both constrained and unconstrained architectures, starting from a PBE clone gives lower errors on WATER27 than starting from random parameters.
In addition, for a fixed initialization strategy, the constrained architecture performs better than the unconstrained one.
For systems chemically related to the training molecules, both forms of physical information---initialization from a stable cloned functional and explicit architectural constraints---improve the outcome of self-consistent optimization.

The conclusions are different for the more demanding Slim16 test.
Here, the optimization on only two small molecules does not generalize well to reaction-energy differences involving unseen chemical environments: the MAD values of the optimized NN-XC models are substantially larger than those obtained with PBE, by roughly an order of magnitude.
At the same time, the total-energy errors remain slightly smaller than those of PBE.
This contrast indicates that the minimal training set can shift the absolute energies toward the CCSD(T) reference, but does not provide enough chemical diversity to improve transferable reaction-energy predictions.

Overall, this proof-of-concept optimization shows that cloned initializations and architectural constraints are beneficial when the test systems remain close to the training domain.
However, training on $\mathrm{H}_2\mathrm{O}$ and $\mathrm{H}_2\mathrm{O}_2$ alone is not sufficient to obtain a broadly transferable functional.
The result supports the role of cloning as a stabilizing initialization strategy, while also showing that chemically diverse correlated reference data are required for systematic improvement beyond PBE.

\subsection{\label{ssec:solid_transfer} Transferability to crystalline solids}

A central question for any learned XC functional is whether it remains meaningful outside the data domain in which it was constructed.
We therefore test transfer from molecular descriptors to crystalline solids, a stringent regime change that is rarely examined explicitly in ML-XC benchmarks.
The cloned networks were trained only on molecular densities, whereas solids sample extended periodic environments, different bonding patterns, and a different electronic-structure implementation.
This test is therefore not a search for improved agreement with experiment, but a diagnostic of whether the learned local representation preserves the PBE fixed point in condensed phases.

We evaluate the NN-XC functionals on a chemically diverse set of crystalline materials spanning several bonding regimes and structural classes.
The set includes simple metals, Ag, Cu, and Na; covalent solids, Si, Ge, and diamond C; binary oxides with polar-covalent and ionic character, SiO$_2$ and MgO; ionic salts, NaCl, LiF, LiCl, KF, CaF$_2$, and MgF$_2$; and layered van der Waals materials, black phosphorus, SnSe, hexagonal BN, and graphite.
Together, these systems probe delocalized metallic densities, directional covalent bonding, ionic and polar environments, and highly anisotropic layered structures.
Initial structures were obtained from the Materials Project~\cite{jain_commentary_2013}.

For each system, the total energy was evaluated as a function of volume by varying the lattice constant around equilibrium.
Eight volume points were used for each equation-of-state curve.
The resulting $E(V)$ data were fitted to the Murnaghan equation of state, yielding the equilibrium lattice constant $a_0$ and bulk modulus $B_0$.
All NN-XC results are compared against PBE calculations performed with the same solid-state setup, so that the errors quantify fidelity to the PBE reference rather than agreement with experiment.
All calculations were performed with the \textsc{siesta} code.
Detailed numerical results and computational settings are provided in Supplementary Sec.~S10.
\begin{figure}[t]
\centering
\includegraphics[width=0.95\linewidth]{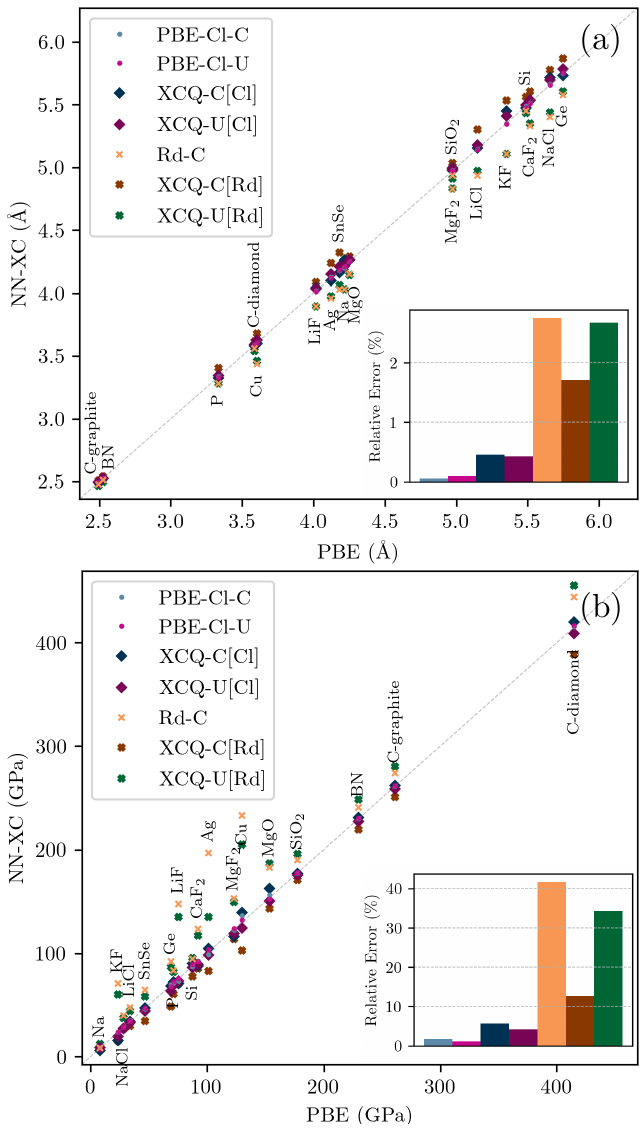}
\caption{Comparison of (a) equilibrium lattice constants $a_0$ and (b) bulk moduli $B_0$ for all solid systems and all NN-XC functionals.
The comparison includes constrained (C) and unconstrained (U) NN-XC models at
three stages: PBE-cloned models (PBE-Cl-C and PBE-Cl-U); self-consistently
optimized models initialized either from the corresponding clone
(XCQ-C[Cl] and XCQ-U[Cl]) or from random parameters
(XCQ-C[Rd] and XCQ-U[Rd]); and untrained random constrained models evaluated directly
without cloning or optimization (Rd-C), where converged calculations
were obtained.
% %
The x-axis shows PBE reference values, the y-axis NN-XC predictions.
The diagonal line marks perfect agreement.
The figures embedded in the lower right edge correspond to the mean relative errors over all the analyzed solids for each network (where color bars match the marker colors of the corresponding network).} 
\label{fig:a0B0}
\end{figure}

Fig. ~\ref{fig:a0B0} summarizes the values for $a_0$ and $B_0$ for the different NN-XC functionals compared to PBE for all the solid systems, as well as the relative errors with respect to PBE.
The most important result is that the cloned NN-XC functionals reproduce PBE structural properties with high accuracy across the entire solid-state test set.
Both constrained and unconstrained PBE clones give equilibrium lattice constants very close to the PBE values, with only small relative deviations.
The agreement is also strong for the bulk moduli, showing that the cloned functionals reproduce not only the equilibrium volumes but also the curvature of the equation-of-state curves.
This demonstrates that a functional cloned from molecular density descriptors can transfer to periodic condensed-matter environments and preserve the local energetics that determine equilibrium structural properties.

The similarity between the constrained and unconstrained cloned models in solids is also informative.
In the molecular tests discussed above, the constrained architecture gave a more accurate PBE clone, especially for total energies.
For the solid-state structural properties considered here, however, both cloned architectures remain close to PBE.
This is consistent with the fact that, at the cloning stage, both networks have learned the same PBE enhancement factors over the descriptor regions most relevant to the solid-state calculations.
The result suggests that the molecular-to-solid transfer of the cloned models is governed primarily by how well the sampled descriptor space overlaps with the descriptors encountered in solids.

An additional and conceptually important result in Fig.~\ref{fig:a0B0} is that Rd-C, the randomly initialized constrained model evaluated without cloning or self-consistent optimization, already yields reasonable equilibrium structural properties for many solids.
This behavior reflects the constrained functional form rather than meaningful information encoded in the random parameters.
The output map bounds the enhancement factors, enforces the uniform-electron-gas limit at $s=0$, and suppresses pathological behavior in the slowly varying density regime that is important for crystalline solids.
This shows that, even before training, the constrained architecture places the neural functional near a physically admissible self-consistent manifold.
This contrasts with unconstrained random networks, which lack such a baseline and are therefore much more prone to unstable or nonphysical behavior.

The self-consistently optimized XCQ models behave differently depending on their initialization.
Models optimized starting from a cloned functional remain comparatively close to the PBE structural predictions, although they no longer reproduce PBE exactly because their parameters have been updated to reduce errors relative to wavefunction-based molecular reference energies.
By contrast, models optimized from random initializations show substantially larger deviations from the PBE solid-state results, and unconstrained randomly initialized networks may fail to converge.
This reinforces the conclusion drawn from the molecular tests: a physically meaningful initialization and, when possible, explicit architectural constraints are important for obtaining stable self-consistent behavior outside the training domain.

Overall, the solid-state evaluation shows that PBE cloning is not merely a molecular interpolation exercise.
The cloned NN-XC functionals transfer to crystalline materials and reproduce PBE equilibrium lattice constants and bulk moduli with high fidelity, despite being trained only on molecular descriptors.
At the same time, the deterioration observed for models optimized from random initializations highlights the sensitivity of self-consistent NN-XC functionals to the quality of the starting point.
This makes the solid-state test an important diagnostic for whether a neural functional has learned a transferable exchange--correlation representation rather than a narrow molecular correction.

\section{\label{sec:conclusions} Discussion and Conclusions}

We have presented XC functional cloning as a diagnostic and initialization strategy for constructing neural XC functional models.
In this approach, a neural model is trained to reproduce a well-known analytic XC approximation on a selected rung of Jacob's ladder; here, we use the GGA functional PBE.
Because the target functional is known, this scheme provides a controlled setting in which the effects of constraints, descriptor sampling, density representation, and self-consistency can be examined separately.
Our main conclusion is that physical constraints shape the learned self-consistent map in ways that go beyond improving a pointwise fit.
Constrained PBE clones reproduce molecular total and energy differences better than otherwise identical unconstrained clones, even though PBE itself already satisfies the imposed exact conditions.
When evaluated in solids, a constrained random model already produces physically reasonable structural predictions before cloning or optimization, showing that the constrained output form acts as a strong architectural prior that places the NN-XC functional near an admissible fixed-point manifold.

We also find that the numerical realization of the density is part of the learning problem.
PBE clones trained on all-electron and pseudopotential densities do not transfer symmetrically across codes, especially for total energies, while reaction-energy errors are more robust because core-related contributions largely cancel, indicating that learned local functionals are tied not only to formal descriptors but also to the density domain sampled by the electronic-structure representation used during training.

Finally, we show that a cloned initialization improves subsequent self-consistent optimization against correlated energies and enables meaningful molecular-to-solid transfer.
The clone PBE functionals from molecular data reproduce PBE lattice constants and bulk moduli across metals, covalent solids, ionic materials, oxides, and layered van der Waals systems, demonstrating that the models are not simple interpolants within the data region they learned from.
At the same time, our results show limited transferability after learning within a small correlated training set. Therefore, systematic improvement beyond PBE will require chemically diverse reference data. This will be studied in a subsequent work.
Together, these results identify constraint-aware cloning as a practical route for building stable, portable, and transferable NN-XC models, and as a broader strategy for developing scientific machine-learning components that must operate inside nonlinear self-consistent solvers.

{\it Acknowledgments} - MVFS, SNR, and KD were funded by the National Science Foundation awards DMR-2427902/2427903. SNR acknowledges the Joan Oró predoctoral grants program of the Department of Recerca i Universitats de la Generalitat de Catalunya and the European Social Fund Plus, 2024 FI-I 00704 and the grant no 2021 SGR 01519 from AGAUR. The ICN2 is funded by the CERCA programme/Generalitat de Catalunya and is supported by the Grant «Excelencia Severo Ochoa» CEX2021-001214-S, funded by MICIU/AEI /10.13039/501100011033. MCG acknowledges support from grant no. PID2024-159869NA-I00 funded by MICIU/AEI/10.13039/501100011033 and ERDF/EU, from grant no. PID2022-140845OB-C66 funded by MCIN/AEI/10.13039/501100011033 and ERDF/EU, from the Diputación Foral de Gipuzkoa through Grants 2024-FELL-000007-01 and 2025-FELL-000009-01, and the technical and human support provided by the DIPC Supercomputing Center. The authors would like to thank Stony Brook Research Computing and Cyberinfrastructure and IACS at SBU for access to the high-performance SeaWulf computing system, funded by the National Science Foundation (awards 1531492 and 2215987) and matching funds from the Empire State Development’s Division of Science, Technology and Innovation (NYSTAR) program (contract C210148).

\bibliographystyle{unsrtnat} % Order by appearance number
\bibliography{__refs}% Produces the bibliography via BibTeX.
\end{document}